\documentclass{amia} 

\usepackage{times}
\usepackage[super]{natbib}
\usepackage[paper=letterpaper, margin=1in]{geometry}
\usepackage[labelfont=bf]{caption}
\usepackage{graphicx}        
\usepackage{xcolor}        

\usepackage{amsmath,amssymb,amsfonts}  
\usepackage{booktabs}        
\usepackage{multirow}        
\usepackage{tabularx}       
\usepackage{algorithm}      
\usepackage{algpseudocode}   
\usepackage{float}    
\usepackage{subcaption} 

\usepackage{pifont} 
\usepackage[hidelinks]{hyperref}

\usepackage{titlesec}
\usepackage[superscript]{cite}
\begin{document}

\title{Magnification-Aware Distillation (MAD): A Self-Supervised Framework for Unified Representation Learning in Gigapixel Whole-Slide Images}

\author{Mahmut S. Gokmen, MS$^1$, Mitchell A. Klusty, BS$^1$, Peter T. Nelson, MD, PhD$^2$, Allison M. Neltner$^2$, Sen-Ching Samson Cheung, PhD$^3$, Thomas M. Pearce, MD, PhD$^4$, David A Gutman, MD, PhD$^5$, Brittany N. Dugger, PhD$^6$, Devavrat S. Bisht, MS$^6$, Margaret E. Flanagan, MD$^7$, V. K. Cody Bumgardner, PhD$^1$ }

\institutes{
    $^1$Center for Applied Artificial Intelligence, University of Kentucky, Lexington, KY\\
    $^2$Sanders-Brown Center on Aging, University of Kentucky, Lexington, KY\\
    $^3$Department of Electrical and Computer Engineering, University of Kentucky, Lexington, KY\\
    $^4$Division of Neuropathology, University of Pittsburgh, Pittsburgh, PA\\
    $^5$Department of Biomedical Informatics, Emory University, Atlanta, GA\\
    $^6$Department of Pathology and Laboratory Medicine, University of California Davis, Sacramento, CA\\
    $^7$Department of Laboratory Medicine and Pathology, University of Texas Health, San Antonio, TX\\
}

\maketitle

\section*{Abstract}

\textit{Whole-slide images (WSIs) contain tissue information distributed across multiple magnification levels, yet most self-supervised methods treat these scales as independent views. This separation prevents models from learning representations that remain stable when resolution changes, a key requirement for practical neuropathology workflows. This study introduces Magnification-Aware Distillation (MAD), a self-supervised strategy that links low-magnification context with spatially aligned high-magnification detail, enabling the model to learn how coarse tissue structure relates to fine cellular patterns. The resulting foundation model, MAD-NP, is trained entirely through this cross-scale correspondence without annotations. A linear classifier trained only on $10\times$ embeddings maintains 96.7\% of its performance when applied to unseen $40\times$ tiles, demonstrating strong resolution-invariant representation learning. Segmentation outputs remain consistent across magnifications, preserving anatomical boundaries and minimizing noise. These results highlight the feasibility of scalable, magnification-robust WSI analysis using a unified embedding space.
}
\section{Introduction}
Computational pathology has emerged as a transformative paradigm in digital diagnostics, leveraging artificial intelligence to analyze gigapixel-scale Whole-Slide Images (WSIs) \cite{Cui2021_1,Niazi2019_2}. These images exhibit a pyramidal, multi-resolution structure that parallels the diagnostic workflow of pathologists, who rely on low-magnification views to understand tissue architecture and switch to high magnification for cellular detail \cite{Acs2020_3, Bera2019_4}. However, this scale-dependent organization poses a central challenge for deep learning: the visual appearance of the same anatomical region changes drastically with magnification, but models must understand how these views relate to effectively utilize the pyramidal structure. Traditional supervised methods \cite{Isensee2020_5, Stringer2020_6, Echle2020_7} operate at fixed resolutions and require extensive pixel-level annotations, limiting scalability.

The recent rise of Self-Supervised Learning (SSL) and Vision Transformers (ViTs) has produced powerful foundation models for pathology \cite{Caron_2021_ICCV_8, Chen_2021_CVPR_9, pmlr-v139-zbontar21a_10}. Most current models, including UNI \cite{Chen2024_11}, UNI2, and Prov-GigaPath \cite{Xu2024_12}, follow a single-scale paradigm and are trained almost exclusively at 20× magnification. While effective, this design prevents them from learning how a 10× contextual view relates to its 40× cellular structure, limiting magnification-consistent representation learning.
More recent developments attempt to incorporate multiple magnifications. Virchow2 \cite{virchow2} increases scale diversity by drawing tiles from $5\times$–$40\times$, yet samples each scale independently, losing the spatial correspondence between coarse and fine detail. PATHS \cite{PATHS} employs a workflow-inspired multi-step patch selection, but its formulation remains oriented toward slide-level classification and relies heavily on external UNI features, creating a substantial preprocessing bottleneck. These approaches demonstrate that simply exposing a model to multiple magnifications is insufficient; current strategies still do not learn how magnification levels relate to one another, preventing coherent, resolution-stable representations. 

To address this gap, we introduce Magnification-Aware Distillation (MAD), a self-supervised framework that extends the teacher–student paradigm \cite{dinov2, Joint_Embedding_Predictive_16, dinov2_register} into a multi-scale training strategy. The teacher network processes low-magnification tiles that provide broad anatomical context, while the student network learns from spatially aligned high-magnification tiles. This design enables the model to internalize how coarse tissue architecture relates to fine-grained cellular patterns \cite{Graham2019_17, Gamper2020PanNukeDE_18}, producing a unified and semantically coherent embedding space across scales. We evaluate this framework on neuropathology WSIs \cite{Srinidhi2021_19, Schmauch2020_20}, a clinically important yet underserved domain in which existing foundation models show limited specialization and insufficient scale-appropriate representation learning \cite{Lu2021_21, Campanella2019_22}.

\noindent \textbf{Contributions.} This work presents three principal contributions to computational pathology. First, we propose a magnification-aware distillation framework that learns semantically linked representations across WSI pyramid levels, the first approach to explicitly encode cross-magnification relationships during self-supervised pre-training. By training a teacher network on low-magnification context (e.g., 10×) to guide a student network processing spatially corresponding high-magnification details (e.g., 40×), the method enables robust coarse-to-fine representation learning that surpasses prior spatial or scale-focused multi-scale approaches \cite{scaling_23}. Second, we demonstrate the clinical utility of this unified embedding space through zero-shot cross-magnification evaluation. A linear classifier trained exclusively on 10× embeddings maintains high performance when applied directly to unseen 40× tiles, whereas conventional models degrade substantially under magnification-induced domain shift \cite{Tellez2019_24}. This capability supports efficient workflows such as low-magnification screening followed by high-resolution analysis without retraining. Third, we present MAD-NP\footnote{\url{huggingface.co/IBI-CAAI/MAD-NP}}, a Vision Transformer foundation model trained with Magnification-Aware Distillation (MAD) for neuropathology. Comprehensive segmentation experiments show that its representations outperform state-of-the-art pathology foundation models (UNI \cite{Chen2024_11}, UNI2\cite{Chen2024_11}, Prov-GigaPath \cite{Xu2024_12}) and a domain-adapted DINOv2 giant baseline trained on identical data, achieving superior clustering quality, cross-magnification consistency, and zero-shot generalization, thereby establishing a new benchmark for magnification-invariant histopathological analysis.

\section{Methods} 
Our methodology establishes a unified pipeline for learning resolution-invariant representations. We first describe the extraction of spatially aligned multi-scale tiles to maintain anatomical correspondence. Next, we detail the Magnification-Aware Distillation (MAD) framework, designed to enforce semantic consistency across magnifications. Finally, we outline the evaluation protocols used to benchmark the model against state-of-the-art approaches.
\subsection*{Multi-Scale Tile Extraction and Indexing}
\begin{figure*}[htbp]
\centering
\includegraphics[width=0.80\linewidth]{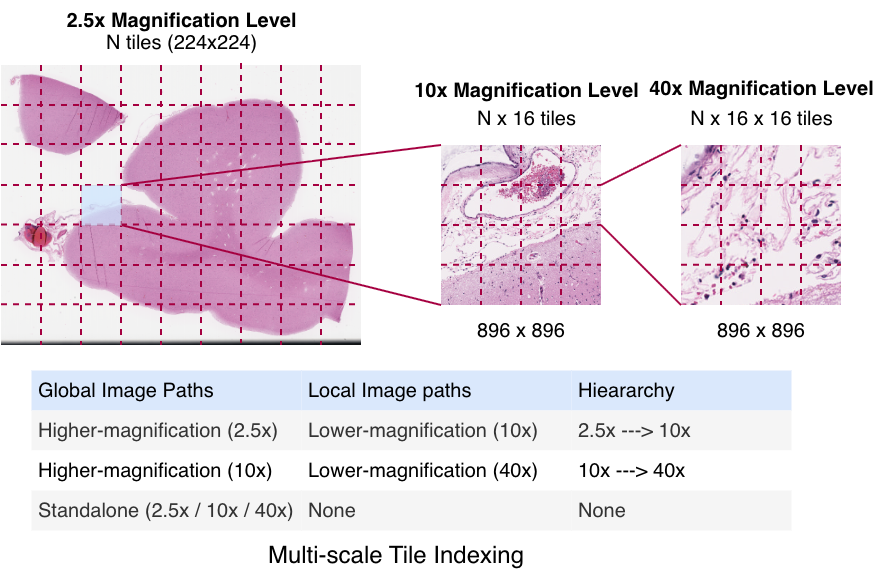}
\caption{\textit{Multi-scale tile generation and indexing across magnification levels. Each $224\times224$ tile at $2.5\times$ corresponds to $256$ spatially aligned tiles at $40\times$ via intermediate $10\times$ tiling.}}
\label{fig:tile_gen}
\end{figure*}

Whole-slide images (WSIs) contain gigapixel-level content, making direct end-to-end processing computationally infeasible. As a result, tiling is the standard strategy for extracting manageable image regions \cite{Cui2021_1}. However, random tiling disregards the spatial correspondence between magnification levels. To address this, we construct multi-scale tiles that preserve anatomical alignment across resolutions, enabling coarse-to-fine representation learning (Figure~\ref{fig:tile_gen}).

Leveraging the intrinsic WSI pyramid, we utilize scanner-generated downsampled layers derived from the $40\times$ native acquisition to access $2.5\times$ and $10\times$ views. The process begins with a $224\times224$ tile at $2.5\times$. This spatially maps to an $896\times896$ region at $10\times$, which is extracted as 16 distinct $224\times224$ tiles. Similarly, each of these $10\times$ tiles projects to a corresponding $896\times896$ region at $40\times$, which is further partitioned into 16 tiles of $224\times224$ pixels. This ensures all model inputs are standardized to $224\times224$, creating a hierarchy where a single low-magnification tile links to 256 spatially aligned high-resolution views.

\textbf{Multi-scale indexing.}
To ensure deterministic correspondence across magnifications, each tile is assigned a hierarchical index. A $2.5\times$ tile at grid position $(i,j)$ maps to 16 tiles at $10\times$ indexed $(i,j,k)$, where $k \in [0,15]$ represents the linearized spatial index within the $4\times4$ sub-grid. Similarly, each $10\times$ tile maps to 16 tiles at $40\times$ indexed $(i,j,k,m)$. This indexing establishes a deterministic multiresolution coordinate system, guaranteeing that teacher and student networks process spatially aligned regions independent of sampling order.
Standalone tiles are also sampled independently at $2.5\times$, $10\times$, and $40\times$, enabling the model to capture both magnification-invariant structure and scale-specific variability.

\subsection*{Magnification-Aware Distillation}

Our training approach is built upon the DINOv2 \cite{dinov2,dinov2_register} self-supervised learning framework, which follows a teacher–student architecture where the student network learns to match the teacher's output through self-distillation. Both models are Vision Transformers (ViTs), and the teacher is updated using an exponential moving average (EMA) of the student's weights to ensure stable and consistent representations. We introduce Magnification-Aware Distillation, which extends this framework through two key modifications: (1) a magnification-aware data augmentation strategy that preserves spatial correspondence across magnification levels while generating diverse views, and (2) an asymmetric teacher-student configuration where the teacher processes low-magnification contextual tiles and the student learns from spatially-aligned high-magnification detail tiles. This design enables the model to capture the  multi-scale relationships between different resolution levels, producing unified representations that maintain semantic coherence across the WSI pyramid.

\textbf{Magnification-Aware View Sampling.}
Standard self-supervised learning (SSL) frameworks typically generate multiple views through stochastic, label-preserving augmentations of a single image. While we retain standard photometric augmentations (e.g., color jittering, solarization) for robustness, our framework introduces a deterministic multi-resolution view sampling strategy that leverages the intrinsic WSI pyramid. This approach is conceptually distinct from classical augmentation; rather than synthetically resizing crops, we extract physically aligned tiles from different magnification levels to construct global and local views.

In this strategy, a global view is sampled from a lower magnification context ($2.5\times$ or $10\times$) and processed by the teacher network. Simultaneously, the student network receives four high-magnification tiles randomly selected from the spatially corresponding region (e.g., 4 out of 16 available sub-tiles). Although the WSI pyramid spans three levels ($2.5\times \to 10\times \to 40\times$), we implement training in a pairwise manner (e.g., $2.5\times \to 10\times$ or $10\times \to 40\times$). This pairwise restriction aligns with the standard two-stream teacher--student architecture of ViTs, allowing the model to bridge the full pyramid through learned transitivity without requiring a computationally expensive multi-tower design.
\begin{table}[H]
\centering
\caption{\textit{Comparison of standard DINO augmentation vs. magnification-aware view sampling.}}
\label{tab:aug_compare}
\resizebox{0.7\textwidth}{!}{%
\begin{tabular}{l|c|c}
\toprule
\textbf{Aspect} & \textbf{Standard DINO} & \textbf{Magnification-Aware (Ours)} \\
\midrule
Input Source & Single image & Multi-scale pairs ($10\times \rightarrow 40\times$) \\
Global Views & 2 augmented crops & 2 augmented context tiles ($10\times$) \\
Local Views & 8 resized crops & 4 native resolution tiles ($40\times$) \\
Total Views per Pass & 10 & 6 (Reduced computational load) \\
Magnification-Awareness & None (Random Resizing) & Explicit via Sampling Path \\
Spatial Alignment & Random & Preserved \\
Detail Preservation & Interpolated / Blurry & Native / High Fidelity \\
\bottomrule
\end{tabular}%
}
\end{table}

Unlike standard DINO, which relies on random cropping and resizing to enforce spatial invariance, our method preserves the native resolution of high-magnification tiles to maintain fine-grained histological structures. To ensure spatial diversity without breaking anatomical correspondence, the local views are not fixed; they are stochastically sampled from the grid of available sub-tiles (selecting 4 out of 16). This acts as a discrete spatial augmentation, forcing the model to correlate variable high-resolution parts with the same global context.

\textbf{Magnification-Aware Distillation Strategy.}
We extend the standard DINOv2\cite{dinov2,dinov2_register} framework with Magnification-Aware Distillation (MAD). Distinct from classical transform-based augmentation, we employ a structural view-based sampling strategy.
In each iteration, the teacher receives a global context tile from a lower magnification level ($L_{low}$), and the student processes high-resolution details from the spatially corresponding next magnification level ($L_{high}$). Specifically, we utilize two transition pairs: $2.5\times \rightarrow 10\times$ and $10\times \rightarrow 40\times$. If the teacher receives a $2.5\times$ tile, the student receives four native-resolution tiles randomly sampled from the corresponding $10\times$ region. Conversely, if the teacher receives a $10\times$ tile, the student learns from the aligned $40\times$ tiles. This pairwise curriculum ensures that the model learns transitive scale relationships across the entire pyramid without requiring simultaneous tri-level inputs.

\begin{figure*}[htbp]
\centering
\includegraphics[width=0.80\linewidth]{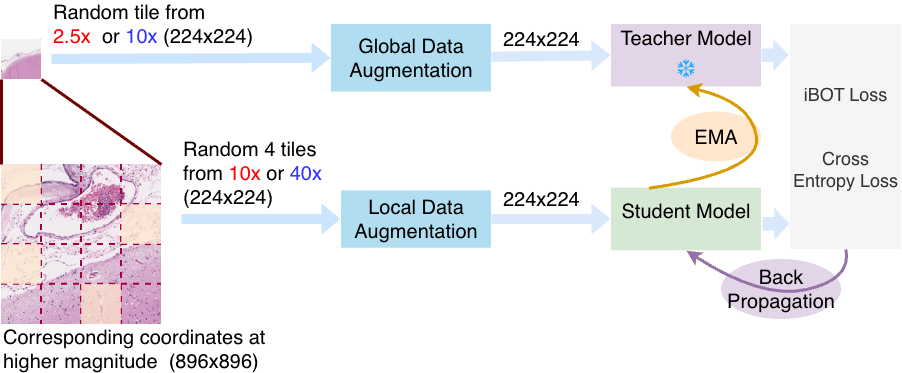}
\caption{\textit{Magnification-Aware Distillation (MAD) training pipeline. Lower-magnification tiles provide global context to the teacher network, while spatially-aligned high-magnification tiles serve as detailed local views for the student network.}}
\label{fig:training}
\end{figure*}
To promote magnification-aware learning, we employ a \textbf{ViT-giant (ViT-g/14) with registers} backbone (40 blocks, 1536 dim, 24 heads, patch size 14, 4 register tokens). The student minimizes cross-entropy loss against the teacher, updated via exponential moving average (EMA). This design introduces student-side tile diversity even with a static global context, improving generalization without new annotations.
\subsection*{Downstream Tasks}

The quality of the learned representations was evaluated using two complementary downstream tasks: tile-level classification and tile-based segmentation. This dual approach facilitates the assessment of both semantic separability and spatial consistency using a frozen ViT backbone. For tile-level classification, each $224 \times 224$ image tile was represented by its corresponding CLS token embedding, and a single-layer linear classifier was trained to map these embeddings to the five definitive tissue categories: \textit{Gray Matter}, \textit{White Matter}, \textit{Leptomeninges}, \textit{Superficial Cortex}, and \textit{Background}. Training was performed using cross-entropy loss and the Adam optimizer with a learning rate of $1\times10^{-4}$ and batch size of 64. This lightweight classifier provides a direct evaluation of feature quality without requiring fine-tuning of the backbone.

For segmentation, the same frozen linear probe trained for classification was utilized. Whole-slide neuropathology images were divided into non-overlapping $224 \times 224$ tiles at $40\times$ and $10\times$ magnifications. Each tile was independently processed by each evaluated model to extract embeddings, and the corresponding class predictions from the linear head were used to reconstruct a region-level segmentation map. This tile-based inference protocol, which generates segmentation maps using tile-level supervision, was applied uniformly across all compared models to ensure consistent evaluation.

\section{Results}
 
This section evaluates the effectiveness of Magnification-Aware Distillation (MAD) across four key dimensions: classification performance, clustering quality, cross-magnification consistency, and zero-shot generalization capabilities. MAD-NP is compared against state-of-the-art foundation models (UNI \cite{Chen2024_11}, UNI2\cite{Chen2024_11}, Prov-GigaPath \cite{Xu2024_12}, Virchow2\cite{virchow2}) and a domain-adapted DINOv2 Giant baseline trained on identical mixed-magnification data using standard DINO self-supervised learning. This controlled comparison isolates the impact of magnification-aware training from data composition or domain adaptation effects. All evaluations use frozen embeddings with a simple linear probe to isolate representation quality from classifier complexity.
\subsection*{Dataset}
The study utilizes a cohort of 61 whole-slide neuropathology images (WSIs) sourced from the UK-ADRC Neuropathology Lab at the University of Kentucky\cite{neuropathlab2025}. The dataset was split into training (50 WSIs) and testing (11 WSIs) sets, ensuring no patient overlap. The slides are annotated with five clinically relevant tissue classes: \textit{Gray Matter}, \textit{White Matter}, \textit{Leptomeninges}, \textit{Superficial Cortex}, and \textit{Background}. Expert neuropathologists performed manual delineation of tissue regions. To ensure label reliability, annotations were validated through a systematic quality control process. Labels for individual tiles were identified by majority voting of pixel classes, acknowledging that tiles may contain mixed tissue boundaries.

\begin{table*}[htbp]
\centering
\caption{\textit{Dataset composition and split. Spatially-aligned tiles preserve multiresolution correspondence. The dataset is split into Training (50 WSIs) and Testing (11 WSIs) sets.}}
\label{tab:tile_magnification_final}
\resizebox{1.0\textwidth}{!}{%
\begin{tabular}{l|cc|c|c||cc}
\toprule
\multirow{2}{*}{\textbf{Tissue Class}}
& \multicolumn{2}{c|}{\textbf{Spatially-Aligned Pairs}}
& \textbf{Standalone}
& \multirow{2}{*}{\textbf{Total Tiles}}
& \multicolumn{2}{c}{\textbf{Data Split}} \\
\cmidrule(lr){2-3} \cmidrule(lr){4-4} \cmidrule(lr){6-7}
& $2.5\times \rightarrow 10\times$ & $10\times \rightarrow 40\times$
& ($2.5 \times$,$10 \times$,$40 \times$)
&
& \textbf{Train} & \textbf{Test} \\
\midrule
Gray Matter        & 1,263 $\to$ 20,208  & 22,460 $\to$ 359,360 & 9,434  & 379,568 & 311,121 & 68,447 \\
White Matter       & 1,283 $\to$ 20,528  & 18,442 $\to$ 295,072 & 7,913  & 315,600 & 258,688 & 56,912 \\
Leptomeninges      & 1,060 $\to$ 16,960  & 12,363 $\to$ 197,808 & 5,356  & 214,768 & 176,039 & 38,729 \\
Superficial Cortex & 786 $\to$ 12,576    & 8,367 $\to$ 133,872  & 3,661  & 146,448 & 120,039 & 26,409 \\
Background         & --                  & --                   & --     & 60,149  & 49,302  & 10,847 \\
\midrule
\textbf{Grand Total} & -- & -- & \textbf{26,364} & \textbf{1,116,533} & \textbf{915,189} & \textbf{201,344} \\
\bottomrule
\end{tabular}
}
\end{table*}

To support multi-scale representation learning and subsequent downstream analyses, each WSI was processed using the multi-scale tile extraction and indexing strategy. This procedure maps low-magnification tiles to their spatially aligned high-magnification regions, creating structured multiresolution sets across $2.5\times$, $10\times$, and $40\times$. Standalone tiles were also sampled independently at each magnification level to ensure comprehensive coverage of scale-specific features. Finally, strict maximum tile counts per class were enforced during the sampling process to ensure a controlled data distribution, as detailed in Table~\ref{tab:tile_magnification_final}.
\subsection*{Classification Performance and Clustering Quality}

The quality and discriminative power of the learned representations were evaluated through classification and clustering experiments using frozen embeddings from all evaluated models. Each $224 \times 224$ tile was encoded into a CLS token embedding. For classification, we employed two protocols across all magnification levels ($2.5\times$, $10\times$, and $40\times$): a standard single-layer linear classifier trained to map embeddings to the five tissue classes, and a non-parametric k-Nearest Neighbors (k-NN, $k=20$) classifier. To ensure statistical reliability, we report the mean performance averaged over five independent runs. Beyond supervised classification, we assessed embedding quality using unsupervised K-Means clustering ($k=5$). Adjusted Mutual Information (AMI) \cite{AMI_paper} quantifies agreement between model-derived clusters and ground-truth labels (higher is better), while the Davies-Bouldin Index (DBI) \cite{DBI_paper} measures cluster compactness and separation (lower is better). Together, these metrics reveal whether embeddings are both semantically coherent and geometrically well-structured.

Beyond supervised classification, we assessed embedding quality using unsupervised K-Means clustering ($k=5$). We computed Adjusted Mutual Information (AMI) to quantify agreement between model-derived clusters and ground-truth labels, and the Davies-Bouldin Index (DBI) to measure cluster compactness. Together, these metrics reveal whether embeddings are both semantically coherent and geometrically well-structured.

\begin{table}[htbp]
\centering
\caption{\textit{Model Performance Comparison. Values represent mean scores averaged over five independent runs. Arrows indicate direction of better performance ($\uparrow$ higher is better, $\downarrow$ lower is better). Best results in bold.}}
\label{tab:model_comparison_updated}
\resizebox{0.75\textwidth}{!}{%
\begin{tabular}{lcccccc}
\toprule
\textbf{Class} & \textbf{MAD-NP} & \textbf{DINOv2 Giant Finetuned} & \textbf{Prov-GigaPath} & \textbf{UNI2} & \textbf{UNI} & \textbf{Virchow2} \\
\midrule
\multicolumn{7}{c}{\textit{Linear F1 Scores} ($\uparrow$)} \\
\midrule
Background & \textbf{0.9188} & 0.9059 & 0.9185 & 0.9148 & 0.9168 & 0.9085 \\
Gray Matter & 0.9328 & 0.9243 & \textbf{0.9339} & 0.9318 & 0.9311 & 0.9230 \\
Leptomeninges & \textbf{0.9127} & 0.8895 & 0.9063 & 0.9045 & 0.9048 & 0.8911 \\
Superficial & \textbf{0.9267} & 0.8964 & 0.9128 & 0.9105 & 0.9089 & 0.8899 \\
White Matter & 0.9625 & 0.9568 & \textbf{0.9652} & 0.9643 & 0.9610 & 0.9551 \\
\midrule
\multicolumn{7}{c}{\textit{k-NN F1 Scores} ($\uparrow$)} \\
\midrule
Background & 0.9160 & 0.8961 & 0.9021 & 0.9052 & \textbf{0.9176} & 0.9018 \\
Gray Matter & 0.9297 & 0.9131 & 0.9311 & 0.9267 & \textbf{0.9359} & 0.9161 \\
Leptomeninges & \textbf{0.9137} & 0.8948 & 0.8947 & 0.8957 & 0.9065 & 0.8883 \\
Superficial & \textbf{0.9196} & 0.8858 & 0.9113 & 0.9116 & 0.9119 & 0.8805 \\
White Matter & 0.9640 & 0.9564 & 0.9674 & 0.9655 & \textbf{0.9703} & 0.9493 \\
\midrule
\multicolumn{7}{c}{\textit{Global Clustering Metrics}} \\
\midrule
Global AMI ($\uparrow$) & \textbf{0.7668} & 0.5275 & 0.4732 & 0.4478 & 0.2975 & 0.3597 \\
Global DBI ($\downarrow$) & \textbf{1.2821} & 1.4133 & 2.0342 & 2.3078 & 1.9559 & 1.2867 \\
\bottomrule
\end{tabular}}
\end{table}

As shown in Table~\ref{tab:model_comparison_updated}, while supervised F1 scores are competitive across models, unsupervised metrics reveal distinct gaps. MAD-NP achieves the highest Global AMI (0.7668) and lowest DBI (1.2821). This divergence suggests that while linear probes can mask embedding irregularities through supervision, MAD-NP possesses a significantly more coherent intrinsic structure. Its magnification-aware training ensures embeddings organize into anatomically meaningful groups, whereas baselines exhibit consistently weaker global organization.

\subsection*{Cross-Magnification Representation Consistency}

A key objective of this study is to determine whether models learn coherent multi-scale representations that maintain consistency across magnification levels. Cross-magnification consistency was measured using spatially-aligned parent–child pairs from the $10\times \rightarrow 40\times$ transition. The $2.5\times \rightarrow 10\times$ transition was excluded from this analysis because the limited number of $2.5\times$ tiles prevented reliable sampling for negative pairs.

Each parent tile at $10\times$ was compared with three categories of $40\times$ child tiles using cosine similarity \cite{cosine_sim}: (1) spatially aligned positive pairs ($S_{pos}$), (2) negative non-aligned pairs of the same tissue class ($S_{neg}^{same}$), and (3) negative pairs from different tissue classes ($S_{neg}^{diff}$). Based on these measures, we define the spatial alignment gap as $\Delta_{hier} = S_{pos} - S_{neg}^{same}$ and the semantic gap as $\Delta_{sem} = S_{neg}^{same} - S_{neg}^{diff}$. Here, $\Delta_{hier}$ measures the model's ability to recognize specific anatomical structures; by subtracting the similarity of same-class textures ($S_{neg}^{same}$) from the exact parent-child match ($S_{pos}$), it shows if the model learns precise spatial location rather than just general texture patterns. Meanwhile, $\Delta_{sem}$ measures how well the model separates different tissue types across magnifications, regardless of spatial alignment.

\begin{table}[htbp]
\centering
\caption{\textit{Cross-magnification consistency for $10\times \rightarrow 40\times$ transition.}}
\label{tab:hierarchical_10x_40x}
\resizebox{1\textwidth}{!}{%
\begin{tabular}{lccccc}
\toprule
\textbf{Model}
& \textbf{Avg Pos. Sim.} ($S_{pos}$)$\uparrow$
& \textbf{Avg Neg. Same} ($S_{neg}^{same}$)$\downarrow$
& \textbf{Avg $\Delta_{hier}$ $\uparrow$ }
& \textbf{Avg Neg. Diff.} ($S_{neg}^{diff}$)$\downarrow$
& \textbf{Avg $\Delta_{sem}$ $\uparrow$ } \\
\midrule
MAD-NP & \textbf{0.716} & 0.577 & \textbf{0.138} & 0.221 & \textbf{0.356} \\
DINOv2 Giant Finetuned & 0.306 & 0.237 & 0.069 & 0.081 & 0.156 \\
UNI2 & 0.488 & 0.417 & 0.071 & 0.343 & 0.074 \\
Prov-GigaPath  & 0.401 & 0.335 & 0.066 & 0.297 & 0.038 \\
UNI  & 0.576 & 0.528 & 0.048 & 0.468 & 0.060 \\
Virchow2 & 0.495 & 0.435 & 0.060 & 0.393 & 0.043 \\
\bottomrule
\end{tabular}
}
\end{table}

The positive similarity values in Table~\ref{tab:hierarchical_10x_40x} illustrate how effectively each model matches a high-magnification tile to its corresponding low-magnification region. Although a $40\times$ tile lacks the global context present at $10\times$, MAD-NP consistently yields higher similarity for aligned pairs. Notably, while exhibiting high negative same-class similarity due to tissue coherence, MAD-NP raises the positive similarity significantly more, resulting in the largest spatial alignment gap ($\Delta_{hier} = 0.138$). While the lack of variance measurements warrants cautious interpretation, this gap is approximately double that of the nearest baseline ($0.069$), suggesting a substantially stronger structural preference for true spatial correspondence beyond mere texture matching. Furthermore, MAD-NP achieves the highest semantic gap ($\Delta_{sem} = 0.356$), ensuring that tiles of different tissues remain well-separated in the feature space even in the absence of spatial alignment.
\begin{figure*}[htbp]
    \centering
    \begin{subfigure}[b]{0.32\textwidth}
        \centering
        \includegraphics[width=\textwidth]{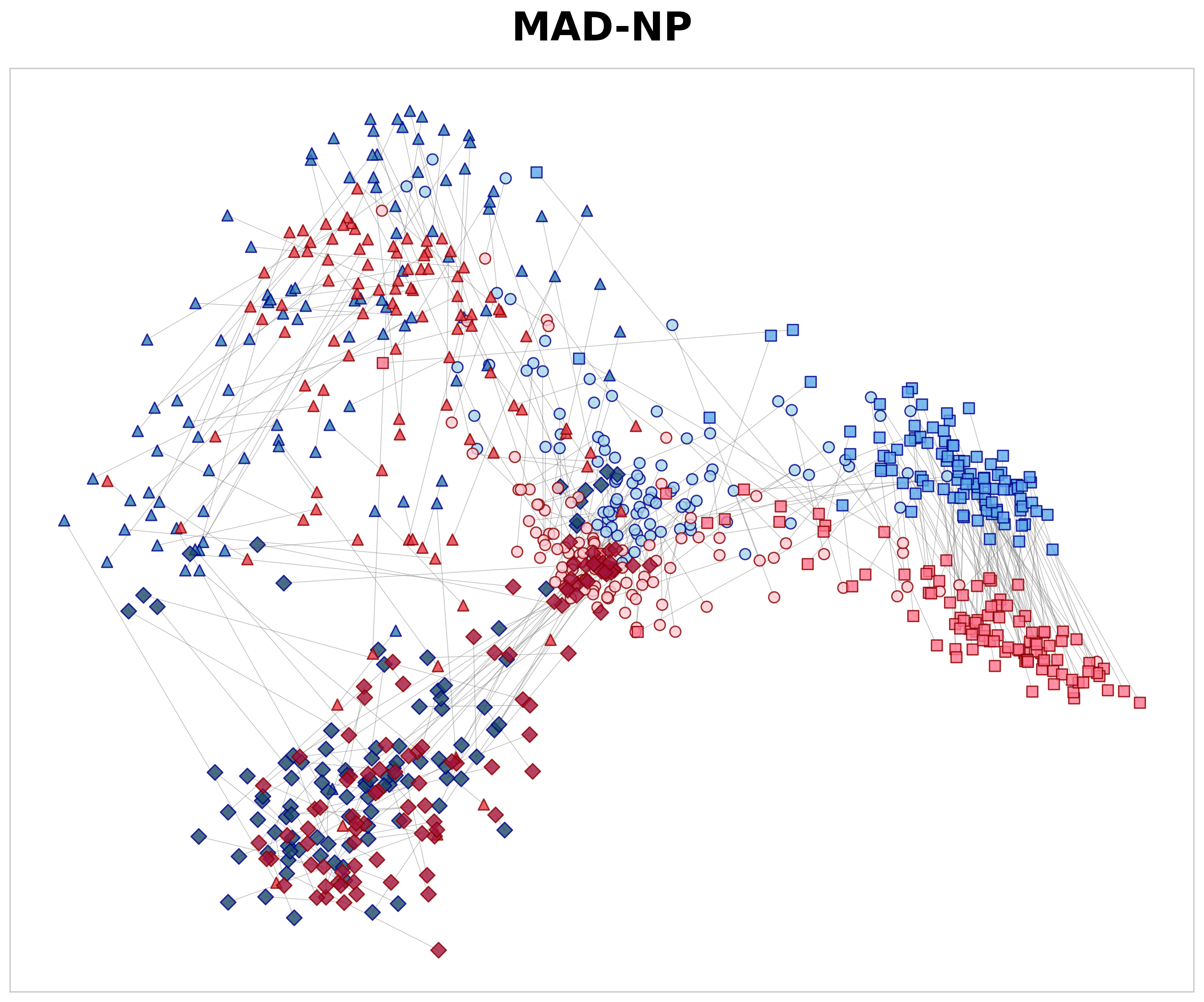}
        \caption{MAD-NP}
        \label{fig:pca_npgiant}
    \end{subfigure}
    \hfill
    \begin{subfigure}[b]{0.32\textwidth}
        \centering
        \includegraphics[width=\textwidth]{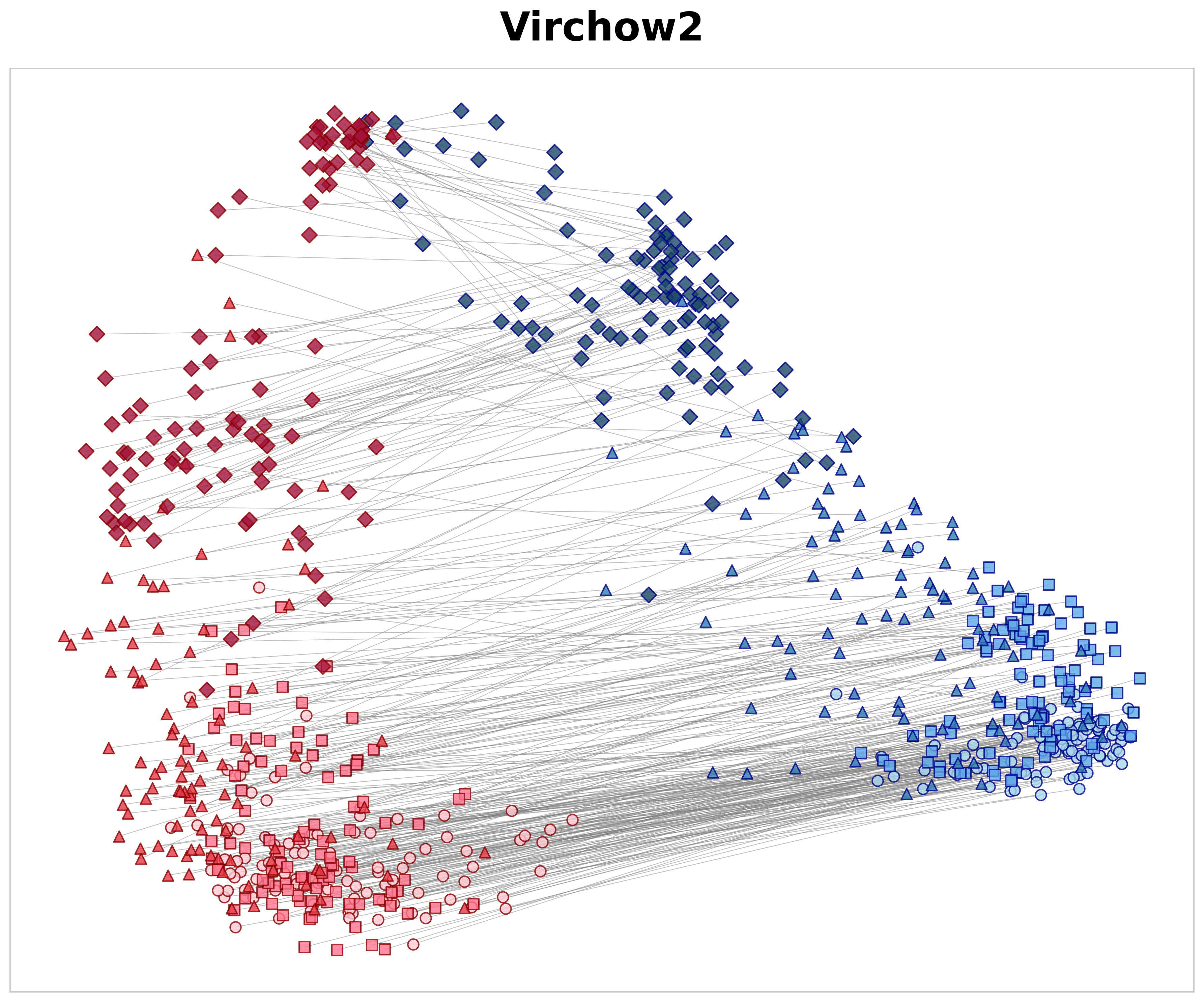}
        \caption{Virchow2}
        \label{fig:pca_virchow2}
    \end{subfigure}
    \hfill
    \begin{subfigure}[b]{0.32\textwidth}
        \centering 
        \includegraphics[width=\textwidth]{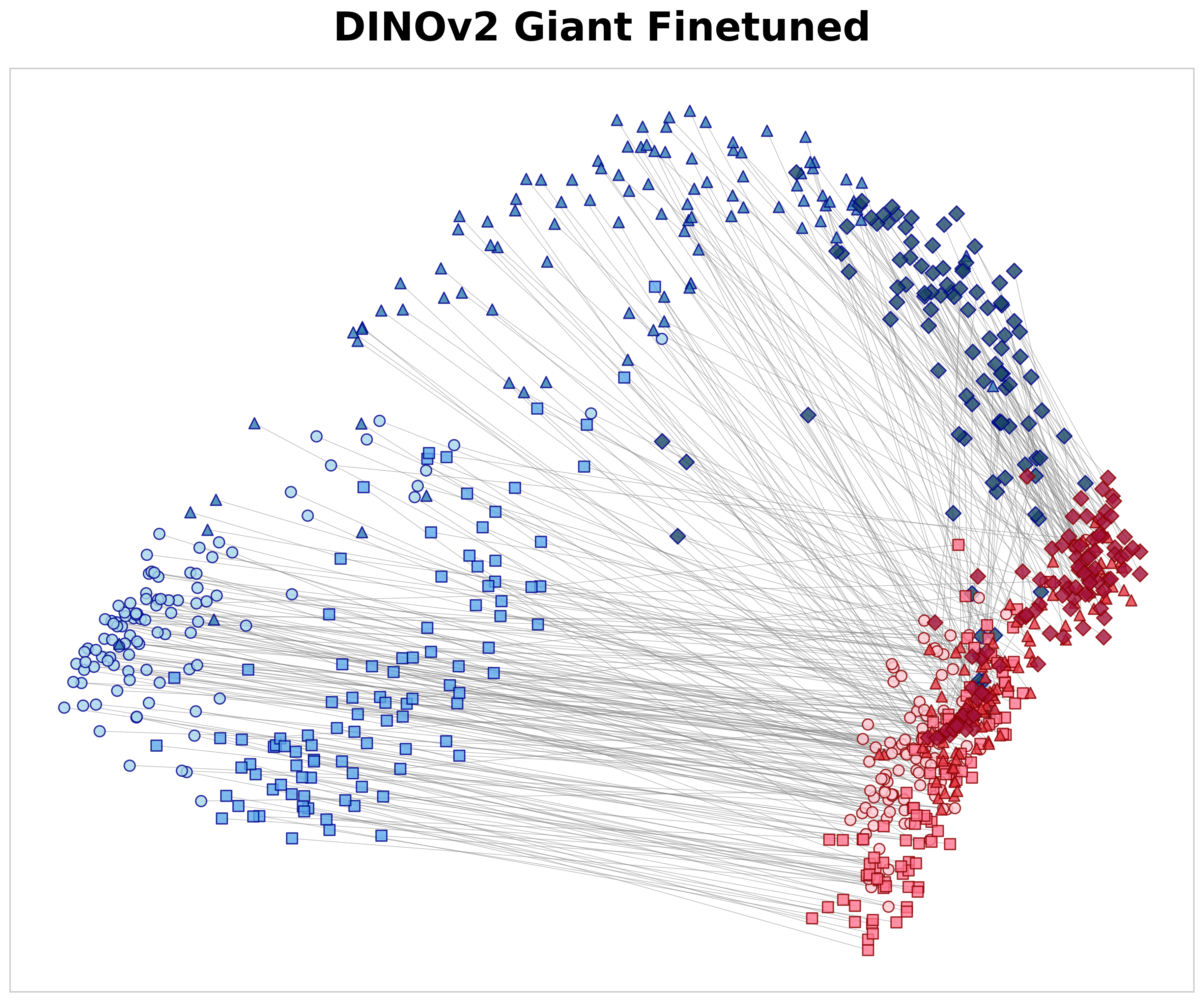}
        \caption{DINOv2 Giant Finetuned}
        \label{fig:pca_dinov2_giant_ft}
    \end{subfigure}
    
\caption{\textit{PCA visualization of 10$\times$ (blue) and 40$\times$ (red) embeddings. Gray lines connect spatially-aligned parent-child pairs. Distinct marker shapes denote the five tissue categories, while variations in color saturation reflect embedding density due to point overlap. MAD-NP produces overlapping clusters where tissue identity is preserved across resolutions, whereas baselines show magnification-dependent separation.}}
    \label{fig:pca_comparison}
\end{figure*}
To visualize these quantitative differences, Figure~\ref{fig:pca_comparison} presents PCA \cite{PCA_paper} projections of $10\times$ and $40\times$ embeddings. As shown in \textbf{Figure~\ref{fig:pca_npgiant}}, MAD-NP embeddings from both levels form overlapping clusters with short connection lines, indicating that spatially aligned parent and child tiles occupy proximal positions in feature space. In contrast, Virchow2 (\textbf{Figure~\ref{fig:pca_virchow2}}) exhibits clear magnification-dependent separation, treating different resolution levels as distinct feature domains. Crucially, this limitation persists even in \textbf{DINOv2 Giant Finetuned} (\textbf{Figure~\ref{fig:pca_dinov2_giant_ft}}), which was trained on the identical multi-scale dataset using the standard DINOv2 training protocol. The model interprets resolution changes as domain shifts, providing further evidence that passive multi-scale exposure is insufficient. Conversely, MAD-NP maintains robust spatial alignment, demonstrating the effectiveness of explicit distillation for unified representation learning.

\subsection*{Cross-Magnification Generalization and Segmentation}

Magnification robustness was evaluated through a cross-magnification generalization experiment. We trained a single linear classifier exclusively on $10\times$ tile embeddings and evaluated it on both $10\times$ (baseline) and unseen $40\times$ tiles. Crucially, no retraining or adaptation was performed for the $40\times$ evaluation. This protocol serves as a strict zero-shot classifier transfer test, assessing whether the representation space remains stable across a four-fold resolution increase. Consistent with previous analyses, the $2.5\times$ level was excluded due to insufficient data for robust training.

\begin{table*}[htbp]
\centering
\caption{\textit{Cross-Magnification Generalization. A linear probe trained only on $10\times$ embeddings is evaluated on both $10\times$ and zero-shot $40\times$ tiles without adaptation.}}
\label{tab:cross_mag_results_expanded}
\resizebox{0.9\textwidth}{!}{%
\begin{tabular}{l|ccc|ccc|c}
\toprule
\multirow{2}{*}{\textbf{Model}} &
\multicolumn{3}{c|}{\textbf{10x Test (Baseline)}} &
\multicolumn{3}{c|}{\textbf{40x Test (Zero-Shot)}} &
\multirow{2}{*}{\textbf{Consistency (\%)}} \\
\cmidrule(lr){2-4} \cmidrule(lr){5-7}
& \textbf{Mean IoU} & \textbf{Mean Dice} & \textbf{Pixel Acc.}
& \textbf{Mean IoU} & \textbf{Mean Dice} & \textbf{Pixel Acc.} & \\
\midrule
\textbf{MAD-NP} & 0.8875 & 0.9368 & 0.9299 &
\textbf{0.8584} & \textbf{0.9193} & \textbf{0.9100} & \textbf{96.7} \\
UNI2 & \textbf{0.8906} & \textbf{0.9388} & \textbf{0.9326} &
0.8018 & 0.8421 & 0.8352 & 90.0 \\
Prov-GigaPath & 0.8879 & 0.9370 & 0.9293 &
0.7245 & 0.7967 & 0.8290 & 81.6 \\
UNI & 0.8961 & 0.9419 & 0.9364 &
0.7212 & 0.7968 & 0.8166 & 80.5 \\
DINOv2 Giant Finetuned & 0.8889 & 0.9375 & 0.9313 &
0.5761 & 0.6477 & 0.6384 & 64.8 \\
Virchow2 & 0.8865 & 0.9332 & 0.9256 & 0.7176 & 0.8233 & 0.7521 & 81.0 \\
\bottomrule
\end{tabular}%
}
\end{table*}

Table~\ref{tab:cross_mag_results_expanded} summarizes the results. On the $10\times$ baseline, all models show comparable performance, suggesting sufficient feature expressiveness for same-scale tasks. However, significant disparities appear in the zero-shot $40\times$ transfer, where the $10\times$-trained classifier is applied directly to high-resolution features. Models trained with standard objectives suffer from domain shift: \textit{DINOv2 Giant Finetuned} exhibits the sharpest decline (Mean IoU dropping from 0.8889 to 0.5761), while UNI and Prov-GigaPath also degrade notably. This indicates that without magnification-aware constraints, high-resolution views are mapped to a distinct feature distribution, breaking the classifier's decision boundary.

In contrast, MAD-NP demonstrates remarkable stability. It retains 96.7\% of its baseline performance, achieving the highest zero-shot metrics at $40\times$ (Mean IoU: 0.8584). This confirms that the model has learned a unified embedding space where $40\times$ structural details map to the same semantic coordinates as their $10\times$ contextual counterparts.

\begin{figure*}[htbp]
\centering
\includegraphics[width=0.8\linewidth]{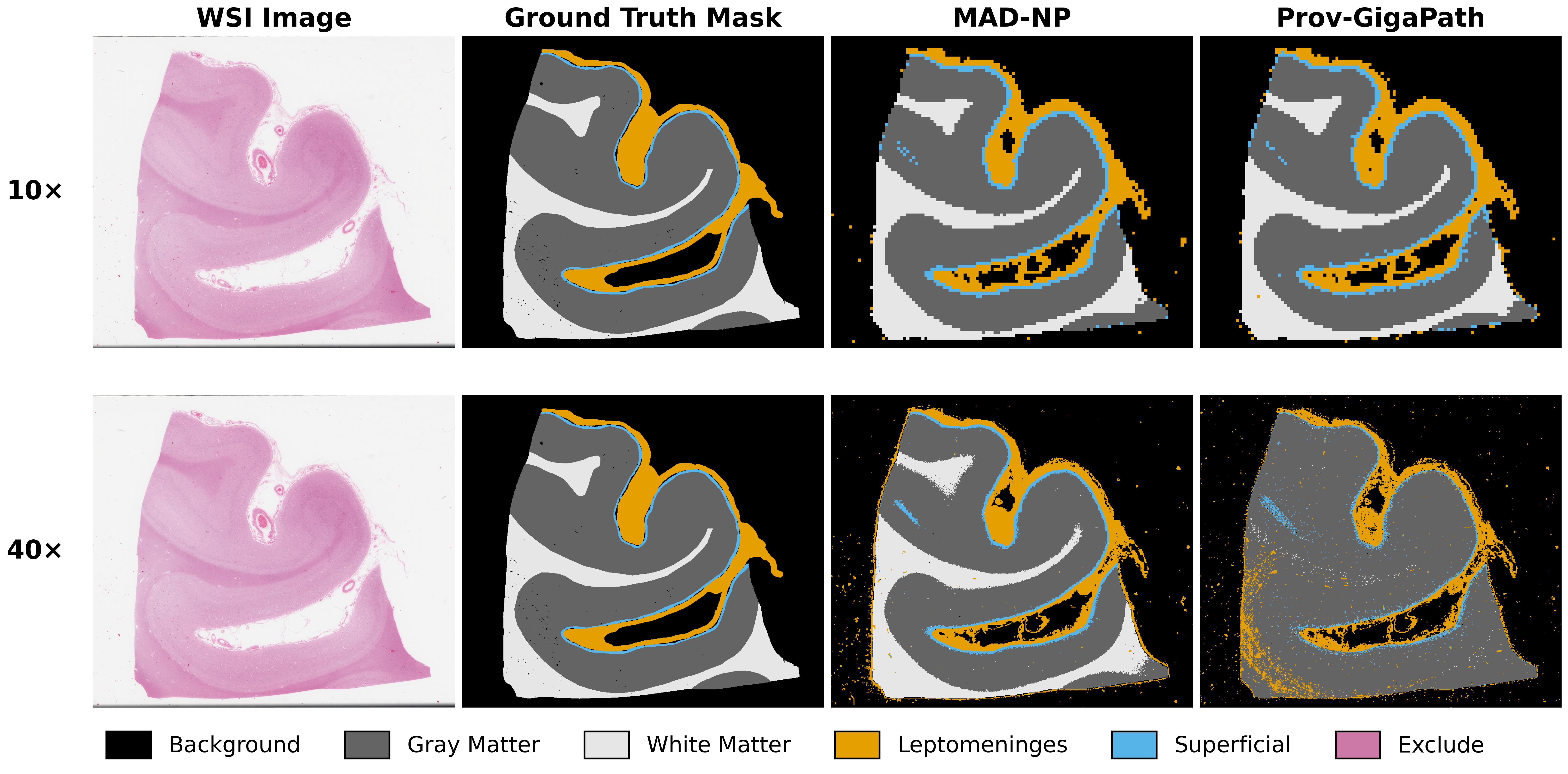}
\caption{\textit{Qualitative segmentation comparison at 10× and 40× magnifications. Columns show: WSI image, ground truth mask, MAD-NP prediction, and Prov-GigaPath prediction.}}
\label{fig:segmentation_comparison}
\end{figure*}

Qualitative results in Figure~\ref{fig:segmentation_comparison} corroborate these findings. MAD-NP generates coherent segmentation maps at both scales, preserving anatomical boundaries and minimizing noise. In contrast, Prov-GigaPath exhibits fragmentation at $40\times$, frequently misclassifying White Matter as Gray Matter. This visual degradation underscores the necessity of explicit alignment for robust cross-magnification analysis.
\section{Discussion and Conclusion}
Magnification-Aware Distillation demonstrates that linking coarse anatomical context with its spatially aligned high-resolution structure provides a viable path for learning unified, multi-scale representations in whole-slide imaging. By assigning low-magnification context to the teacher and high-magnification detail to the student, the framework models how tissue architecture and cellular morphology jointly define neuropathological patterns. This design yields a stable embedding space that remains consistent when resolution changes, enabling tasks such as low-magnification training with high-magnification inference.
Our findings challenge the prevailing assumption in computational pathology that Vision Transformers can implicitly learn global anatomical patterns solely from small, high-resolution patches. Instead, our results suggest that explicitly providing aligned low-magnification context significantly stabilizes the embedding space and improves representational coherence. Furthermore, this framework offers a promising avenue for leveraging archival datasets scanned at lower magnifications (e.g., $20\times$). By training models to correlate available lower-magnification views with high-magnification details, MAD could effectively enhance the utility of historical archives, enabling robust analysis even when modern $40\times$ scans are unavailable.

The strong agreement between 10$\times$-trained classifiers and zero-shot 40$\times$ predictions, together with robust segmentation performance, indicates that compositional cross-scale relationships can be internalized effectively without additional annotations. While the dataset contains a relatively limited number of $2.5\times$ regions, expanding this portion would allow more complete assessment of cross-scale transfer. Additionally, while validated on neuropathology slides, broader validation across diverse histological domains is necessary to confirm generalizability. Overall, the results highlight the value of magnification-aware supervision for creating magnification-robust foundation models and illustrate a promising direction for scalable computational pathology pipelines that operate reliably across the full resolution range of WSIs.
 
\section{Acknowledgments}
This research was supported in part by the National Institutes of Health under award numbers P30AG072946 and U24NS133945. The content is solely the responsibility of the authors and does not necessarily represent the official views of the NIH.

\makeatletter
\renewcommand{\@biblabel}[1]{\hfill #1.}
\makeatother

\bibliographystyle{vancouver}
\bibliography{amia}  

\end{document}